\documentclass{aastex7}

\usepackage{hyperref}

\newcommand{\ldl}{$\lambda/\Delta\lambda$}
\newcommand{\teff}{$T_{\rm eff}$}
\newcommand{\logg}{$\log{g}$}
\newcommand{\Lbol}{$L_\mathrm{bol}$}
\newcommand{\kzz}{$\log{\kappa_{zz}}$}

\newcommand{\vsini}{$v\sin{i}$}
\newcommand{\loglbol}{$\log_{10}{L_{\mathrm{bol}}/L_\odot}$}

\newcommand{\mjup}{M$_{Jup}$}
\newcommand{\rjup}{R$_{Jup}$}

\begin{document}

\title{JWST Observations of the Metal-poor T Dwarf WISEA J155349.96+693355.2: \\ The First Brown Dwarf Associated with the Gaia-Enceladus Milky Way Substructure}

\author[0000-0002-1125-7384]{Aaron M. Meisner}
\affiliation{NSF National Optical-Infrared Astronomy Research Laboratory, 950 N. Cherry Ave., Tucson, AZ 85719, USA}
\affiliation{Center for Astrophysics $|$ Harvard \& Smithsonian, 60 Garden St., Cambridge, MA 02138, USA}
\affiliation{Radcliffe Institute for Advanced Study at Harvard University, 10 Garden Street, Cambridge, MA 02138, USA}
\email{aaron.meisner@noirlab.edu}

\author[0000-0002-6523-9536]{Adam J. Burgasser}
\affiliation{Department of Astronomy \& Astrophysics, UC San Diego, La Jolla, CA 92039, USA}
\email{aburgasser@ucsd.edu}

\author[0000-0002-5370-7494]{Chih-Chun Hsu}
\affiliation{Center for Interdisciplinary Exploration and Research in Astrophysics (CIERA), Northwestern University, 1800 Sherman Ave,
Evanston, IL, 60201, USA}
\email{chsu@northwestern.edu}

\author[0000-0003-0548-0093]{Sherelyn Alejandro Merchan}
\affiliation{Department of Astrophysics, American Museum of Natural History, Central Park West at 79th St., New York, NY 10024, USA}
\affiliation{Department of Physics, Graduate Center, City University of New York, New York, NY, USA}
\email{sherelyna12@gmail.com}

\author[0000-0001-6251-0573]{Jacqueline K. Faherty}
\affiliation{Department of Astrophysics, American Museum of Natural History, Central Park West at 79th St., New York, NY 10024, USA}
\email{jfaherty@amnh.org}

\author[0000-0002-2011-4924]{Genaro Su\'arez}
\affiliation{Department of Astrophysics, American Museum of Natural History, Central Park West at 79th St., New York, NY 10024, USA}
\email{gsuarez@amnh.org}

\author[0000-0003-3047-607X]{ZengHua Zhang}
\affiliation{School of Astronomy and Space Science, Nanjing University, 163 Xianlin Avenue, Nanjing 210023, China}
\affiliation{Key Laboratory of Modern Astronomy and Astrophysics, Nanjing University, Ministry of Education, Nanjing 210023, China}
\email{zz@nju.edu.cn}

\author[0000-0003-0398-639X]{Roman Gerasimov}
\affiliation{Department of Physics and Astronomy, University of Notre Dame, Nieuwland Science Hall, Notre Dame, 46556, Indiana, USA}
\email{rgerasim@nd.edu}

\author[0000-0003-2094-9128]{Christian Aganze}
\affiliation{Kavli Institute for Particle Astrophysics \& Cosmology, Stanford University, Stanford, CA 94305, USA}
\email{caganze@stanford.edu}

\author[0000-0002-3612-8968]{Nicolas Lodieu}
\affiliation{Instituto de Astrof\'isica de Canarias (IAC), Calle V\'ia L\'actea s/n, E-38200 La Laguna, Tenerife, Spain}
\affiliation{Departamento de Astrof\'isica, Universidad de La Laguna (ULL), E-38206 La Laguna, Tenerife, Spain}
\email{nlodieu@iac.es}

\author[0000-0003-4636-6676]{Eileen C. Gonzales}
\affiliation{Department of Physics and Astronomy, San Francisco State University, 1600 Holloway Avenue, San Francisco, CA 94132, USA}
\email{egonzales@sfsu.edu}

\author[0000-0002-6294-5937]{Adam C. Schneider}
\affil{United States Naval Observatory, Flagstaff Station, 10391 West Naval Observatory Rd., Flagstaff, AZ 86005, USA}
\email{aschneid10@gmail.com}

\author[0000-0003-3050-8203]{Stanimir A. Metchev}
\affiliation{Department of Physics and Astronomy, The University of Western Ontario, 1151 Richmond St, London, ON N6A 3K7, Canada}
\email{smetchev@uwo.ca}

\author[0000-0001-7896-5791]{Dan Caselden}
\affiliation{Department of Astrophysics, American Museum of Natural History, Central Park West at 79th St., New York, NY 10024, USA}
\email{dancaselden@gmail.com}

\author[0000-0001-7780-3352]{Michael C. Cushing}
\affiliation{Ritter Astrophysical Research Center, Department of Physics \& Astronomy, University of Toledo, 2801 W. Bancroft St., Toledo, OH 43606, USA}
\email{Michael.Cushing@utoledo.edu}

\author[0000-0002-9807-5435]{Christopher A. Theissen}
\affiliation{Department of Astronomy \& Astrophysics, UC San Diego, La Jolla, CA 92039, USA}
\email{ctheissen@ucsd.edu}

\author[0000-0003-0774-6502]{Jason J. Wang}
\affiliation{Center for Interdisciplinary Exploration and Research in Astrophysics (CIERA), Northwestern University, 1800 Sherman Ave,
Evanston, IL, 60201, USA}
\email{jason.wang@northwestern.edu}

\author[0009-0009-4489-0192]{Harshil Kothari} 
\email{harshil177@icloud.com}
\affiliation{Ritter Astrophysical Research Center, Department of Physics \& Astronomy, University of Toledo, 2801 W. Bancroft St., Toledo, OH 43606, USA}

\begin{abstract}

We present JWST NIRSpec and MIRI observations of the metal-poor T dwarf WISEA J155349.96+693355.2.
The combined NIRSpec/prism plus MIRI/LRS  spectrum (R $\sim 100$) provides 0.6--12~$\mu$m spectroscopic coverage that reveals the presence of H$_2$O, CH$_4$, and NH$_3$ absorption features,
enhanced collision-induced H$_2$ absorption, and weak or absent CO and CO$_2$ absorption,
all indicators of a metal-poor low-temperature atmosphere. 
Comparison of the spectral data to a suite of atmosphere models yields an
effective temperature $T_{\rm eff}$ = 960$^{+28}_{-32}$~K and a solar-scaled metallicity of [M/H] = $-1.16^{+0.28}_{-0.25}$~dex.
We find indications that the uncertain  trigonometric parallax, likely subject to Lutz-Kelker bias, may be overestimated. Adopting the  photometric distance yields 
a bolometric luminosity {\loglbol} = $-$5.29 $\pm$ 0.10, from which we infer $T_{\rm eff}$ = $974 \pm 56$~K, consistent with the temperature derived via our atmospheric model fits.
From the moderate-resolution (R $\sim 3,000$) NIRSpec/G395H spectrum, we measure a radial velocity of $-163 \pm 5$~km~s$^{-1}$, which combined with the source's measured proper motion and estimated distance
indicates a highly eccentric ($e$ $\geq$ 0.9) Galactic orbit consistent with an accreted halo object.
The $UVW$ kinematics of WISEA J155349.96+693355.2, its bulk metallicity, and its modest alpha enrichment ([$\alpha/\textrm{Fe}$] = +0.15$^{+0.04}_{-0.06}$~dex) are all equivalent to stars identified in
the Gaia-Enceladus Milky Way substructure, making it the first brown dwarf with kinematic and chemical evidence of association.
\end{abstract}


\section{Introduction} 

An ever-more detailed picture of the scale, structure, and evolution of the Milky Way and its satellite systems has emerged through deep and wide-scale optical surveys such as the Sloan Digital Sky Survey \citep[SDSS;][]{York_SDSS}, the Panoramic Survey Telescope and Rapid Response System \citep[Pan-STARRS;][]{Chambers_2016}, Gaia \citep{Gaia_canonical}, the Dark Energy Camera Legacy Survey \citep[DECaLS;][]{Dey_2019}, and the Dark Energy Survey \citep[DES;][]{DES}, among others.
These surveys have identified and characterized stellar members of the Galaxy's thick disk, halo, bulge, and stellar streams \citep[e.g.,][]{Juric_2008, Newberg_2002, Yanny_2003, Bernard_2014, Helmi_2018}. 
However, they are less sensitive to the Galaxy's brown dwarf population (M $\lesssim 0.075$~M$_\odot$; \citealt{1962AJ.....67S.579K,1963ApJ...137.1121K,1963PThPh..30..460H}), abundant and long-lived sources which carry their own insights about the kinematic and chemical evolution of the Milky Way \citep{Dino_smart_paper}, as well as environmental influences on low-mass star formation \citep[e.g.,][]{Kroupa_2001,Luhman_review,Yan_2024,Bate_2025}.
Present ground-based surveys are generally limited to identifying low-temperature brown dwarfs in the immediate solar neighborhood, making it difficult to assemble large enough samples of substellar members of the Galaxy's thick disk or halo for population studies \citep{Zhang_2019,Dino_smart_paper,Burgasser_2025}.

Deeper and wider near-infrared space-based surveys by JWST \citep{Gardner_2006}, Euclid \citep{Euclid_Q1}, and the Roman Space Telescope \citep{Spergel_2015}
promise to greatly enhance the volume over which we can study cool (sub)stellar objects, including thick disk and halo brown dwarfs \citep[e.g.,][]{2021MNRAS.501..281S,Honaker_2024,Kiwy_2025,ZJ_Zhang_2025}. 
Indeed, deep JWST surveys focused on extragalactic populations have already uncovered low-temperature brown dwarfs at kiloparsec distances, well beyond the Milky Way's thin disk (e.g., \citealt{Burgasser_2024,2024ApJ...975...31H,2025ApJS..281...49T,2025arXiv251101167M,2025arXiv251000111H,2026MNRAS.tmp..248L}).
Potential insights from these deeper samples include mapping age- and mass-dependent features in the the Milky Way's thin disk scale height \citep{Ryan_2017,Aganze_2022a,Aganze_2022b,Holwerda_2024}, 
metallicity dependencies of the low-mass mass function and star-brown dwarf mass limit \citep[e.g.,][]{2011ApJ...736...47B,Gerasimov_2024a,Gerasimov_2024b}, 
new approaches to age-dating old clusters and Galactic populations from the stellar-substellar luminosity gap \citep{2004ApJS..155..191B,Gerasimov_2024b},
and insights on cold atmosphere chemistry through variations in elemental abundances \citep[e.g.,][]{Faherty_silane,Burgasser_PH3}.

Our ability to study the broader Milky Way with cold brown dwarfs hinges on accurate inference of physical parameters from deep spectrophotometry, which can be achieved by calibrating atmosphere and evolutionary models on nearby brown dwarfs
with low metallicities, unusual abundances, and/or kinematics indicative of membership in the thick disk or halo.
One such object is WISEA J155349.96+693355.2 \citep[hereafter J1553;][]{Meisner_2020b}, a mid-T subdwarf discovered by the Backyard Worlds: Planet 9 citizen science project \citep{Kuchner_2017}. 
Initial analysis of near-infrared photometry, spectroscopy, and astrometry of this source by \cite{Meisner_2021} indicated a subsolar metallicity ([M/H] $\approx -0.5$~dex) and potentially extreme kinematics ($\mu \approx 2.15''$/yr, $V_{tan} \approx 400$~km/s). 
Further analysis of the spectral data by \citet{Burgasser_2025} indicated an effective temperature {\teff} $\approx$ 1200~K, [M/H] $\approx$ $-$1~dex, and an extreme but uncertain radial velocity (RV) of +110 $\pm$ 90 km/s, suggesting membership in the Helmi streams \citep{Helmi_1999}.

Here, we present JWST NIRSpec and MIRI spectroscopic observations of J1553 which extend this object's spectroscopic coverage to 0.6--12~$\mu$m. These data provide improved determinations of its temperature, metallicity, and surface gravity, and a precise and accurate measurement of its RV that indicates membership in the Gaia-Enceladus Milky Way substructure \citep{Belokurov_2018,Helmi_2018,Myeong_2018}.
In $\S$\ref{sec:obs} we describe the JWST observations and reductions for J1553. In $\S$\ref{sec:analysis} we present analysis of these data, including 
measurement of J1553's bolometric luminosity, 
fitting of the low-resolution spectra to several grid models to infer atmosphere parameters, 
and measurement of its RV to assess kinematics. 
We discuss these results in $\S$\ref{sec:discussion},
providing kinematic and chemical evidence of J1553's membership in Gaia-Enceladus.
We summarize our findings in $\S$\ref{sec:conclusion}.

\section{JWST Observations \& Reductions} \label{sec:obs}

J1553 was observed with JWST/NIRSpec \citep{Jakobsen_2022} and JWST/MIRI \citep{Rieke_2015}
as part of program JWST-GO-04668 (PI: Burgasser). On 2024 November 18 (UT), the source was observed with NIRSpec using its prism grating, S200A1 0$\farcs$2 fixed slit, and CLEAR filter,
which provided low-resolution spectra ({\ldl} = 50--300) spanning 0.6--5.2~$\mu$m.  
After acquisition with the F110W broad-band filter ($\lambda = 1.0$-1.3~$\mu$m) and SUB2048 subarray, the source was observed in spectral mode at two spatial nod positions along the slit with a sub-pixel dither (0.5 pixels = 0$\farcs$05). Two exposures
per position were acquired, each consisting of 120 sample-up-the-ramp (SUTR) groups with 1.6~seconds integrations per group, for a total exposure time of 748~seconds. 
On 2024 November 21 (UT), the source was observed with NIRSpec using its G395H grating, S200A1 plus S200A2 slit combination, and F290LP filter, which provided moderate-resolution spectra ({\ldl} = 2,000--3,700) spanning 2.9--5.1~$\mu$m.
The dual slit enabled coverage of the 3.7--3.8~$\mu$m gap between NIRSpec's NRS1 and NRS2 detectors.
After acquisition with the F110W filter, J1553 was observed in two spatial nod positions for each slit,
with 60 SUTR groups and 5.5~seconds integrations per group, for a total effective exposure time of 1319~seconds.
On 2025 February 5 (UT), the source was observed with MIRI using its Low-Resolution Spectrometer (LRS) double-prism dispersion mode,  4{\farcs}7 $\times$ 0{\farcs}51 single slit, and P750L filter, 
which provided low-resolution spectra ({\ldl} = 40--160) spanning 5--14~$\mu$m.  
After acquisition with the F1000W (10~$\mu$m) broad-band filter, the source was observed in spectral mode at two spatial nod positions along the slit separated by 1$\farcs$9, with one exposure
per position consisting of 100 SUTR groups with 2.8~seconds integrations per group, for a total exposure time of 555~seconds. 

For all three datasets, we analyzed the pipeline data reduction products 
provided by the standard JWST science calibration pipeline version 1.16.1 \citep{Bushouse_2024} and served through the Barbara A.\ Mikulski Archive for Space Telescopes (MAST). The Calibration Reference Data System context files used were jwst\_1293.pmap for NIRSpec and jwst\_1303.pmap for MIRI.
Reduction included background, dark current, and bias subtraction; flat field, gain scale, and linearity corrections; bad pixel flagging; count rate extraction; saturation correction; and 1D boxcar spectral extraction. 
The final 1D spectra have median signal-to-noise (S/N) per pixel values of 213 for the NIRSpec/prism data at 4.1~$\mu$m, 74 for the NIRSpec/G395H data at 4.1~$\mu$m,
and 46 for the MIRI/LRS data at 8.5~$\mu$m.

\begin{figure*}
\plotone{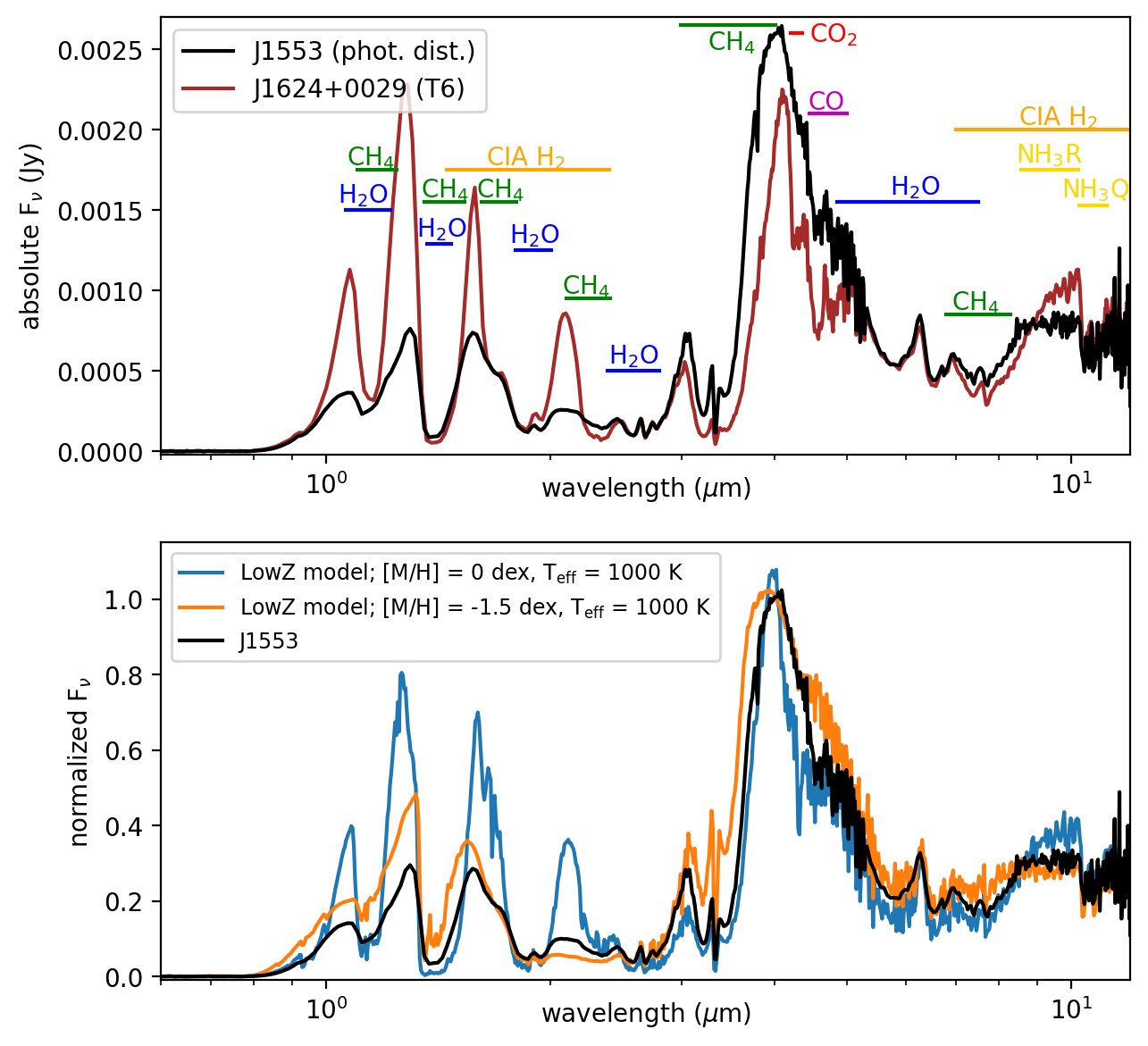}
\caption{Top: Combined 0.6--12~$\mu$m low-resolution spectrum of J1553 from JWST NIRSpec/prism and MIRI/LRS observations.
The J1553 spectrum (black) has been calibrated to absolute $F_\nu$ units using Spitzer ch2 photometry 
and its photometric distance from \citealt{Meisner_2020b}. For comparison, we show the low-resolution JWST spectrum of the roughly solar metallicity T6 dwarf J1624
(brown; data from \citealt{Beiler_2024}), also scaled to absolute fluxes. Approximate wavelength ranges of strong absorption features from H$_2$O, CH$_4$, H$_2$ CIA, CO$_2$, CO, and NH$_3$ are labeled (see also the absorption cross-sections displayed in the top panel of Figure \ref{fig:SED_chemistry_context}) Bottom: Comparison of the J1553 low-resolution spectrum against two \teff = 1000~K LowZ models \citep{Meisner_2021}, one model with solar metallicity (blue) and the other model with metallicity of $-1.5$~dex (orange). Aside from metallicity, these two LowZ models both have identical parameters: \logg = 5.0, \kzz = 2, and C/O = 0.1. The spectra in this panel are all normalized to have mean F$_{\nu}$ of unity between 3.9~$\mu$m and 4.1~$\mu$m.
\label{fig:NIRSpec_and_MIRI}}
\end{figure*}

Figure~\ref{fig:NIRSpec_and_MIRI} displays the combined low-resolution JWST NIRSpec/prism plus MIRI/LRS spectrum. 
The spectra were stitched together after 
scaling the MIRI/LRS spectrum by a factor of 1.038 to align it with the NIRSpec/prism spectrum in the overlapping 4.9--5.1~$\mu$m region.
Although the MIRI/LRS red limit is nominally 14~$\mu$m, low S/N in science frames and certain reference/calibration files curtails the usable wavelength range in practice\footnote{\url{https://jwst-docs.stsci.edu/known-issues/miri-known-issues/miri-lrs-known-issues}}.
The combined JWST NIRSpec+MIRI spectrum captures 95.6\% of J1553's total luminosity, based on the analysis presented in
$\S$\ref{sec:lbol}.

\section{Analysis} \label{sec:analysis}

\subsection{Temperature from Bolometric Luminosity} \label{sec:lbol}

Bolometric luminosity ($L_{\rm bol}$) is a fundamental observable for brown dwarf atmospheres
when a distance and sufficiently sampled spectral energy distribution (SED) are available \citep[e.g.,][]{2015ApJ...810..158F,Beiler_2024,Sanghi_2023}. 
To compute $L_{\rm bol}$, we used the absolute flux-calibrated NIRSpec/G395H spectrum extended shortward to 0.6~$\mu$m with the NIRSpec/prism spectrum and longward to 12~$\mu$m with the MIRI/LRS spectrum as the input SED 
for SEDkit \citep[][see \citealt{Alejandro_2025} for further details]{SEDkit}.
The spectra were calibrated to the absolute Spitzer ch2 (4.5~$\mu$m) magnitude using $m_{ch2}$ = $15.458 \pm 0.018$ mag (Vega) based on the photometry from \cite{Meisner_2021} and the photometric distance of $39^{+5}_{-4}$~pc from \cite{Meisner_2020b}, yielding $M_{ch2}$ = $12.50^{+0.22}_{-0.28}$. We prefer this photometric distance over the more uncertain trigonometric parallax distance from \citet[][26$^{+7}_{-4}$~pc]{Zhang_2025} for reasons discussed in $\S$\ref{sec:distance}.
No additional photometry was used in computing $L_{\rm bol}$ as there are currently no measurements beyond the
spectral range provided by the JWST spectroscopy. SEDkit extends the SED beyond the observed spectrum by appending an appropriately scaled Rayleigh-Jeans tail to approximate the 12~$\mu$m $< \lambda <$ 1000~$\mu$m emission.
Integrating over the absolute flux-calibrated spectrum 
yields {\loglbol} = $-$5.29$\pm$0.10
where the uncertainty accounts for the 12\% uncertainty in the photometric distance.

We computed the effective temperature using the Stefan-Boltzmann relation, assuming a radius of $0.77 \pm 0.01$~R$_{Jup}$ based on the evolutionary models of \cite{Saumon_2008} for an age estimate of 10 $\pm$ 3 Gyr, motivated both by J1553's likely membership in the Galactic halo (see $\S$\ref{sec:kinematics_discussion}) and candidate membership in the  Gaia-Enceladus Milky Way substructure ($\S$\ref{sec:discussion}). The latter estimate is based on the range of Gaia-Enceladus star ages reported in
\cite{Feuillet_2021}.
We used the cloud-free, solar-metallicity evolutionary model from \cite{Saumon_2008}, as this is the only evolutionary model available within SEDkit that extends to ages $> 10$~Gyr. 
We note that the Sonora Bobcat \citep{Sonora_bobcat} cloud-free evolutionary models, which include tracks extending to ages $>$ 10~Gyr at a subsolar metallicity of $-0.5$~dex, yield a radius of $0.76 \pm 0.02$~R$_{Jup}$,  consistent with the radius derived from the \cite{Saumon_2008} models. 
This lends confidence that our temperature estimate is
not strongly affected by the choice of evolutionary model grid.

As a cross-check, we performed an alternative integration of the J1553 SED combining the observed Spitzer photometry \citep{Meisner_2020b} with JWST spectroscopy. The resulting bolometric luminosity agrees with our JWST-only \Lbol~to within $< 1$\%. We do not attempt any SED integration with the WISE photometry, as J1553 is subject to blending at WISE's $\sim$6$''$ angular resolution \citep{Meisner_2021}.

From {\loglbol} = $-$5.29$\pm$0.10, we inferred $T_{\rm eff}$ = $974 \pm 56$~K, where we have combined uncertainties from \Lbol~and the adopted radius.
This value is significantly lower than the {\teff} = 1225$^{+70}_{-75}$~K inferred from 
prior model fitting of J1553's near-infrared spectrum \citep{Burgasser_2025},
although that fit had no distance constraint. On the other hand, it matches well with the $T_{\rm eff} = 960^{+28}_{-32}$~K derived from our NIRSpec+MIRI spectral model fits presented in $\S$\ref{sec:grid}.
For completeness, we also re-ran our \Lbol~analysis using J1553's 
trigonometric distance from \citet[][26$^{+7}_{-4}$~pc]{Zhang_2025},  yielding {\loglbol} = $-$5.66 $\pm$ 0.22 and $T_{\rm eff}$ = 785$^{+101}_{-99}$~K, the larger uncertainties arising from the large ($\pm$21\%) uncertainty in the trigonometric parallax. 
We argue against this lower luminosity and temperature in $\S$\ref{sec:distance}.

\subsubsection{Mass Estimate from Bolometric Luminosity}

Using the \cite{Saumon_2008} evolutionary models within SEDkit and an age of 10 $\pm$ 3 Gyr, J1553's luminosity corresponds to a mass of 
$67^{+2}_{-4}$~\mjup~based on its photometric distance and
$58^{+7}_{-10}$~\mjup~based on its parallax distance.
Both of these mass estimates are in the substellar regime.

\subsection{On the Distance to J1553} \label{sec:distance}

\begin{figure*}[ht!]
\plotone{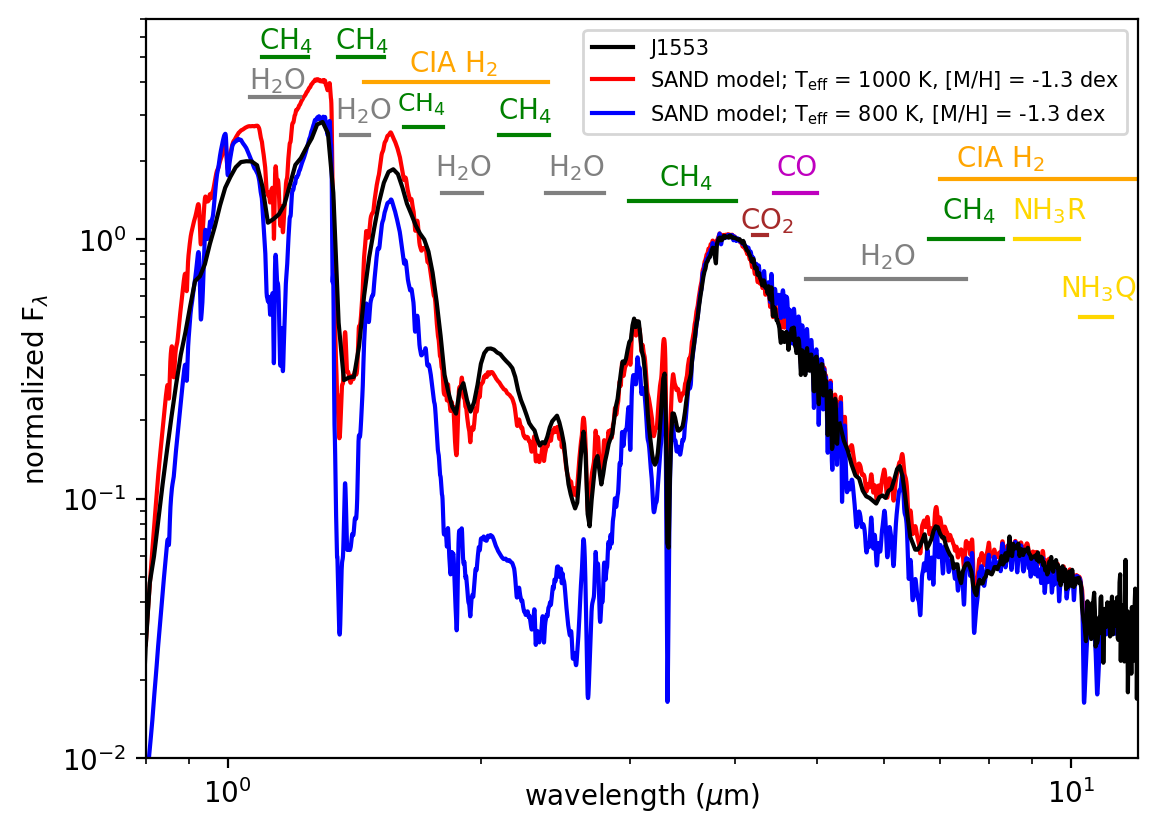}
\caption{Comparison of the J1553 low-resolution JWST spectrum (black) against two models from the SAND grid \citep{Alvarado_2024}. The SAND model shown in red has \teff = 1000~K,
the model in blue has \teff = 800~K, and all other parameters are identical ([$\alpha$/Fe] = 0.15~dex, [M/H] = $-$1.3~dex, and \logg = 4.5; see $\S$\ref{sec:grid}).
All spectra have been normalized between 3.9~$\mu$m and 4.1~$\mu$m. Relevant wavelength ranges of strong molecular absorption are labeled. The \teff = 1000~K model provides a better overall fit to the spectral morphology.
\label{fig:hot_cold_models}}
\end{figure*}

The distinct temperatures inferred from our bolometric luminosity analysis for photometric ($T_{\rm eff}$ = $974 \pm 56$~K) and trigonometric ($T_{\rm eff}$ = 785$^{+101}_{-99}$~K) distances merits scrutiny. 
Figure \ref{fig:hot_cold_models} compares the J1553 NIRSpec+MIRI spectrum to two metal-poor ([M/H] = $-$1.3~dex) SAND atmosphere grid models \citep{Alvarado_2024}:
one with \teff = 1000~K aligned with the photometric distance,
and one with \teff = 800~K aligned with the trigonometric distance,
with all other parameters being equal.
Normalizing these spectra at 4~$\mu$m, we see that the 1000~K model provides a much better match to the overall spectrum of J1553.
The 800~K model more accurately reproduces the peak fluxes at 1.0 and 1.25~$\mu$m, but
significantly underpredicts the observed flux 
in the major H$_2$O and CH$_4$ absorption bands 
at 1.35~$\mu$m $< \lambda < $ 3.5~$\mu$m and 5.4~$\mu$m $< \lambda < $ 6.8~$\mu$m, and also in the NH$_3$ band near 10~$\mu$m.

Furthermore,
our model fitting analysis in $\S$\ref{sec:grid} yields a temperature more consistent with the photometric distance ($T_{\rm eff} = 960^{+28}_{-32}$~K). This model fitting analysis also yields a radius of 0.78$\pm$0.10~\rjup~when scaling the J1553 spectrum to 
the photometric distance, in good agreement with the evolutionary model radius of 0.77$\pm$0.01~\rjup. 
Scaling to the trigonometric distance at the same temperature requires a nonphysically small radius of $0.50 \pm 0.11$~\rjup. The agreement in both parameters suggests that the true J1553 distance is more in line with the photometric distance estimate.

An overestimated trigonometric parallax (underestimated distance) can be understood by accounting for Lutz-Kelker bias \citep{1973PASP...85..573L}, a well-known phenomenon whereby low-S/N trigonometric parallaxes tend to be overestimated.
The J1553 photometric distance has a central value of 39~pc, $\sim 1.5\times$ larger than the 26~pc distance corresponding to the \cite{Zhang_2025} central parallax value. Estimates for the mean Lutz-Kelker absolute magnitude correction for a 20\% error in parallax in a uniformly distributed stellar sample are typically $-$0.2~mag to $-$0.5~mag (10\%--26\% increase in distance), but can exceed $-$1~mag (60\% increase in distance; \citealt{1998MNRAS.294L..41O,2003MNRAS.338..891S}), making a 1.5$\times$ distance correction plausible given the low S/N parallax.

Overall, we favor the hypothesis that J1553's distance is closer to $\sim 39$~pc, with a corresponding $T_{\rm eff} \approx 950$~K and {\loglbol} = $-$5.29$\pm$0.10.
For the remainder of this paper, we adopt the photometric distance ($39^{+5}_{-4}$~pc) for all analyses and discussion.

\subsection{Spectral Features}
\label{sec:spectral_features}

In Figure \ref{fig:NIRSpec_and_MIRI} (top panel), we compare the J1553 spectrum to 
the low-resolution JWST NIRSpec/prism and MIRI/LRS spectrum of the roughly solar metallicity T6 dwarf SDSS J162414.37+002915.6 (hereafter J1624; \citealt{Strauss_1999,Beiler_2024}), similarly stitched and calibrated to absolute fluxes using J1624's ch2 apparent mag and trigonometric parallax \citep{Kirkpatrick_2021a}.
J1624 has an effective temperature (\teff = 987$^{+22}_{-60}$~K; \citealt{Beiler_2024}) similar to that
derived for J1553 (see $\S$\ref{sec:grid}), a surface gravity of {\logg} = 4.89$^{+0.08}_{-0.05}$,
and a metallicity [M/H] = $-$0.18$^{+0.03}_{-0.02}$~dex \citep{Kothari_2026}. J1624 is thus a representative example of the local field T dwarf population
based on its 
near-solar metallicity, field surface gravity, thin disk kinematics, and lack of association with any young moving group \citep{Dino_smart_paper}.
J1624 has also been designated as the near-infrared spectral standard for type T6 \citep{Burgasser_2006}.

Both objects' spectra are strongly shaped by H$_2$O and CH$_4$ absorption that lead to prominent peaks at 1.05~$\mu$m, 1.25~$\mu$m, 1.6~$\mu$m, 
and 4--5$~\mu$m.
The first three peaks are less sharp in J1553's spectrum as compared to J1624, with J1553's 1.6~$\mu$m peak showing a gradual ramp-down toward longer wavelengths reminiscent of structure seen in the extreme T-type subdwarfs WISEA J041451.67$-$585456.7 and WISEA J181006.18$-$101000.5 \citep{Schneider_2020,Lodieu_2022,Zhang_W1810}.
J1624 also shows peaks at 2.1~$\mu$m and 9--12~$\mu$m which are flattened in the spectrum of J1553. 
The different brightnesses of these peaks can be attributed to 
strong H$_2$ collision-induced absorption \citep[CIA;][]{1969ApJ...156..989L,Saumon_2012} in the spectrum of J1553, a common feature of metal-poor ultracool dwarf atmospheres \citep[e.g.,][]{2003ApJ...592.1186B,2011MNRAS.414..575M,Lodieu_2019}.
The flattening of J1553's spectrum at 9--12~$\mu$m is rare among the overall mid-T dwarf population; for instance, none of the mid-T dwarfs in the Spitzer Infrared Spectrograph library \citep{IRS_library} exhibit such flattening around 9~$\mu$m.
We also see that the 4--5~$\mu$m peak is broader in J1553 as compared to J1624. The blue side of this peak is shaped by CH$_4$ absorption which appears weaker in J1553. The red side of this peak is shaped by deep troughs from CO$_2$ and CO absorption in J1624, whereas J1553 shows minimal if any CO$_2$ absorption and very weak CO absorption. 
Again, the relative lack of CO and CO$_2$ in the spectrum of J1553 is consistent with a very metal-poor atmosphere \citep[e.g.,][]{Faherty_silane,Burgasser_PH3}. 
NH$_3$ absorption near 10~$\mu$m is prominent for J1624 and present but weaker in J1553, likely suppressed by strong CIA H$_2$ absorption.
The distinct spectral features of J1553 compared to J1624 across the 0.6--12~$\mu$m band corroborate its status as a low-temperature, metal-poor object.

The bottom panel of Figure \ref{fig:NIRSpec_and_MIRI} compares J1553's low-resolution spectrum to two \teff = 1000~K LowZ models \citep{Meisner_2021}, one with solar metallicity and the other with [M/H] = $-1.5$~dex. The [M/H] = $-1.5$~dex model is a far better match to J1553's spectrum than the solar metallicity model. In particular, the strong flattening of J1553's K-band peak, the relative faintness of J1553's Y, J, and H peaks, J1553's relatively round 4-5~$\mu$m peak, and the muted impact of NH$_3$ absorption near 10~$\mu$m are all features qualitatively well-captured by the [M/H] = $-1.5$~dex model but very poorly matched by the solar metallicity model. This again provides compelling evidence of J1553's subsolar metallicity.

\begin{figure*}
\includegraphics[width=7.10in]{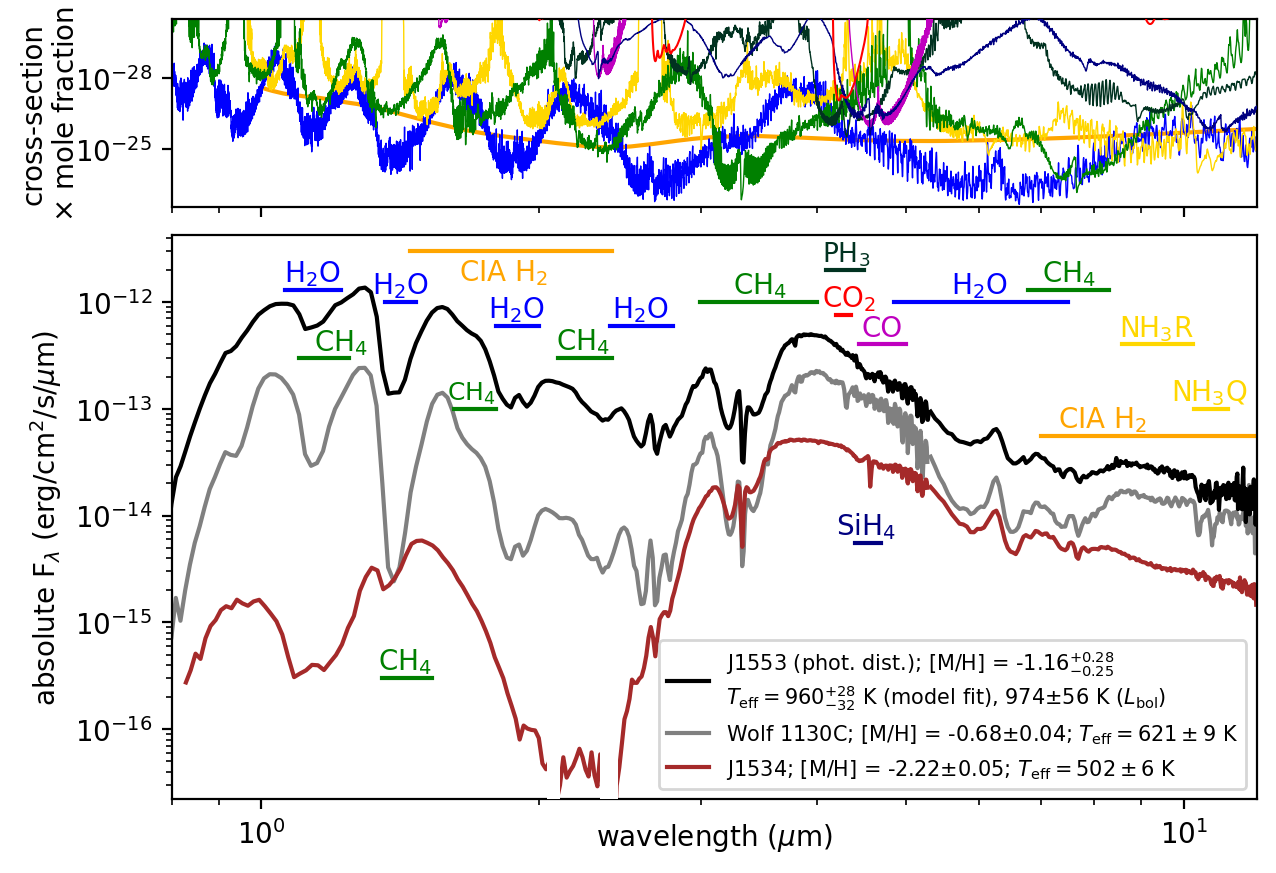}
\caption[Absolute SEDs of J1553, Wolf 1130C, and J1534]{Bottom: Absolute SEDs of J1553 (black line),
Wolf~1130C \citep[gray line;][]{Mace_2013b,Burgasser_PH3},
and J1534 \citep[brown line;][]{Faherty_silane}.
For the latter two sources, $T_{\rm eff}$ values in the annotations have been derived from {\Lbol} measurements \citep{Burgasser_PH3,Faherty_silane}. For J1553, the legend lists both our \teff~derived from SAND model fits ($\S$\ref{sec:grid}), and from \Lbol~using the photometric distance. We show the J1553 spectrum here scaled to absolute flux using its photometric distance.
Metallicities for Wolf 1130C and J1534 are from retrieval analysis \citep{Burgasser_PH3,Faherty_silane}. Approximate wavelength ranges of strong absorption features from H$_2$O, CH$_4$, H$_2$ CIA, PH$_3$, SiH$_4$, CO$_2$, CO, and NH$_3$ are also labeled. Top: Absorption cross-section in cm$^2$/molecule multiplied by mole fraction, for each molecule labeled in the bottom panel, with corresponding colors in the bottom and top panels. The cross-sections were obtained from ExoMol \citep{Tennyson_2016} for SiH$_4$ and 
HITRAN \citep{Gordon_2022} for all other species, and evaluated at $T = 750$~K and $P = 1$~bar. The mole fractions for H$_2$O, CH$_4$, PH$_3$, CO$_2$, CO, and NH$_3$ are from \cite{Burgasser_PH3}, appropriate to Wolf 1130C, with the CO$_2$ mole fraction corresponding to its upper limit value. The mole fraction for SiH$_4$ is from \cite{Faherty_silane}, appropriate for J1534.
\label{fig:SED_chemistry_context}}
\end{figure*}

Figure \ref{fig:SED_chemistry_context} compares J1553's low-resolution JWST spectral energy distribution (SED) to equivalent data for the metal-poor T/Y dwarfs
Wolf~1130C \citep{Mace_2018,Burgasser_PH3} and WISEA J153429.75$-$104303.3 \citep[hereafter, J1534;][]{Kirkpatrick_2021b,Meisner_2020a,Faherty_silane}.
Many of the peculiarities seen in the spectrum of J1553 are also found in these sources, notably broad 4--5~$\mu$m peaks with weak or absent CO and CO$_2$, 
and a flattened 9--12~$\mu$m peak with weakened NH$_3$.
In J1553 we do not detect clear features from phosphine (PH$_3$) or silane (SiH$_4$), present in the spectra of Wolf~1130C and J1534, respectively.
Based on the SEDs in Figure \ref{fig:SED_chemistry_context}, J1553 appears warmer (brighter) than Wolf~1130C ($T_{\rm eff} = 621 \pm 9$~K; \citealt{Burgasser_PH3})
and J1534 ($T_{\rm eff} = 502 \pm 6$~K; \citealt{Faherty_silane}), while the strength of its CIA H$_2$ absorption appears to be intermediate between these sources, suggesting a metallicity between that of Wolf 1130C ([M/H] = $-0.68 \pm 0.04$~dex; \citealt{Burgasser_PH3}) and J1534 ([M/H] = $-2.22 \pm 0.05$~dex; \citealt{Faherty_silane}).

\subsection{Grid Model Fits} \label{sec:grid}

To improve our estimate of {\teff}, and infer surface gravity ({\logg} in units of cm/s$^2$), metallicity ([M/H]), and other parameters,
we compared the combined NIRSpec/prism and MIRI/LRS absolute flux-calibrated spectrum 
to several pre-calculated forward model grids. 
We defer a more extensive retrieval analysis \citep[e.g.,][]{Line_2014,Gonzalez_2021,Hood_2024,Burgasser_PH3}  to a future study.

We considered seven atmosphere model grids which encompass the estimated temperature and metallicity of J1553 from prior model fits and our luminosity analysis above. These models include:\footnote{A subset of these models were downloaded from the Spanish Virtual Observatory (SVO) Theory Server, available at \url{https://svo2.cab.inta-csic.es/theory/newov2}.}
ATMO++ (250~K $\leq$ {\teff} $\leq$ 1200~K, $-$1.0~dex $\leq$ [M/H] $\leq$ +0.3~dex; \citealt{Phillips_2020,2023AJ....166...57M});
BT-Dusty  (1000~K $\leq$ {\teff} $\leq$ 4000~K, $-$2.5~dex $\leq$ [M/H] $\leq$ +0.5~dex; \citealt{2012RSPTA.370.2765A});
Sonora Elf Owl (500~K $\leq$ {\teff} $\leq$ 2400~K, $-$1.0~dex $\leq$ [M/H] $\leq$ +1.0~dex), including 
the original model set \citep{Mukherjee_2024} and sets that include vertically-mixed PH$_3$ \citep{2024ApJ...973...60B} and enhanced CO$_2$ \citep{2025RNAAS...9..108W};
LowZ (500~K $\leq$ {\teff} $\leq$ 1600~K, $-$2.5~dex $\leq$ [M/H] $\leq$ +1.0~dex; \citealt{Meisner_2021});
and SAND (700~K $\leq$ {\teff} $\leq$ 4000~K, $-$2.4~dex $\leq$ [M/H] $\leq$ +0.3~dex; \citealt{Alvarado_2024,Gerasimov_2024a}).
All fits were conducted using the \texttt{ucdmcmc} package version 1.4.2 \citep{ucdmcmc}.
The spectrum of J1553 and atmosphere models were interpolated onto a common wavelength scale based on the observed data that accounts for the varying resolution of both NIRSpec/prism and MIRI/LRS spectra, and evaluated in absolute $F_\lambda$ flux density units.
For each model set, 
an initial grid search was conducted to identify the single best-fitting model spectrum,
using a modified reduced chi-square statistic to evaluate the quality of fit:
\begin{equation}\label{eqn:chi}
    X^2 \equiv \frac{1}{DOF}\sum\limits_{i=1}^N\frac{(O[\lambda_i]-\alpha{M[\lambda_i]})^2}{\sigma^2[\lambda_i]},
\end{equation}
where
$O[\lambda_i]$ is the observed flux density;
$M[\lambda_i]$ is the model flux density;
$\sigma[\lambda_i]$ is the observed flux density uncertainties;
$\alpha$ is a scaling factor that minimizes $X^2$
\citep{2008ApJ...678.1372C}, 
and DOF is the degrees of freedom, equal to the number of data points $N$ minus the number of model parameters including $\alpha$ (DOF = 853 or 854 depending on the model set).
The sum is conducted over the spectral range 0.8~$\mu$m $\leq$ $\lambda$ $\leq$ 12~$\mu$m.
Note that the scale factor $\alpha = \left({R}/10~\mathrm{pc}\right)^2$ provides an estimate of the source radius.

\begin{figure*}[ht!]
\centering
\includegraphics[width=0.75\textwidth]{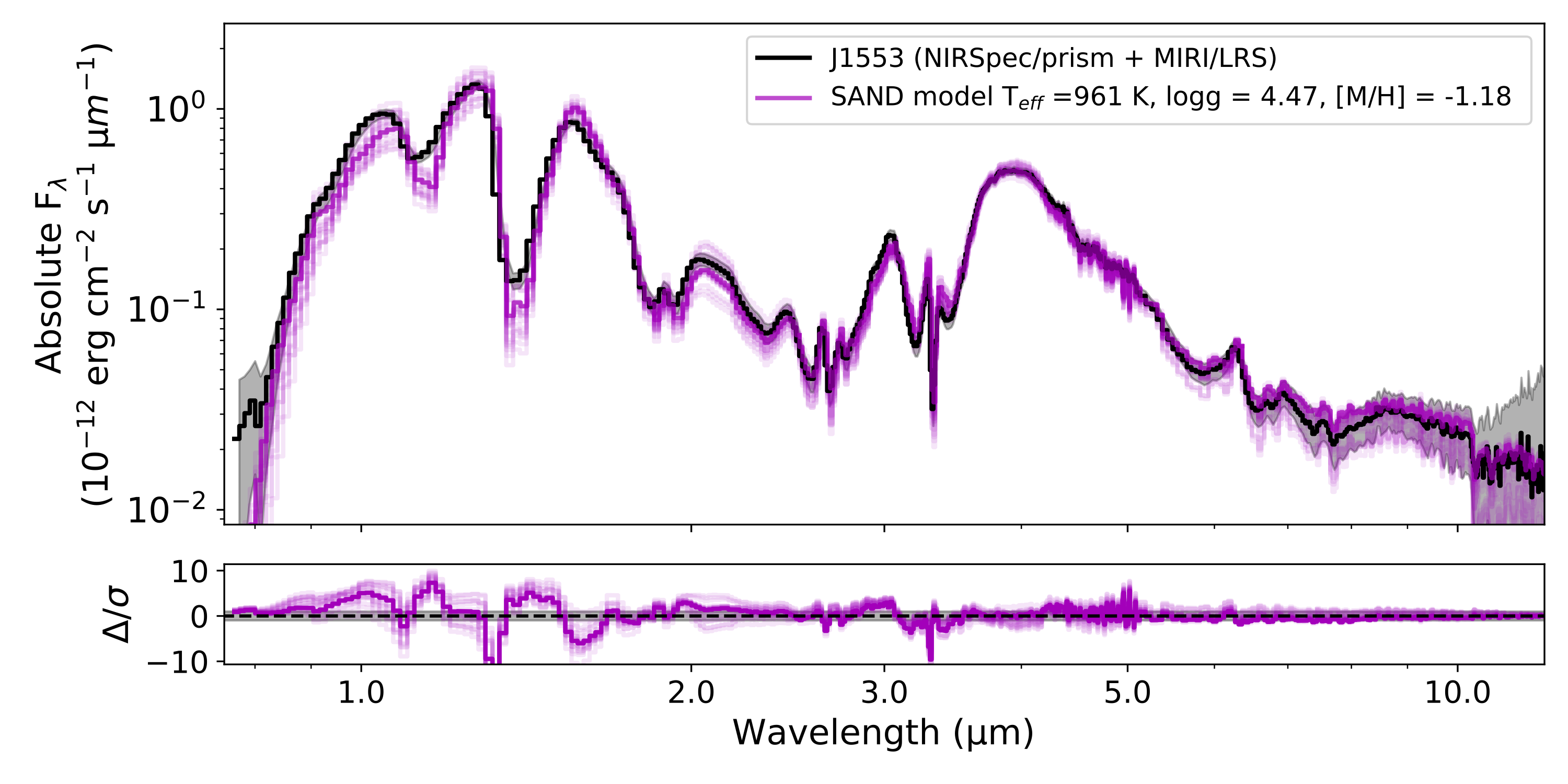} \\
\includegraphics[width=0.75\textwidth]{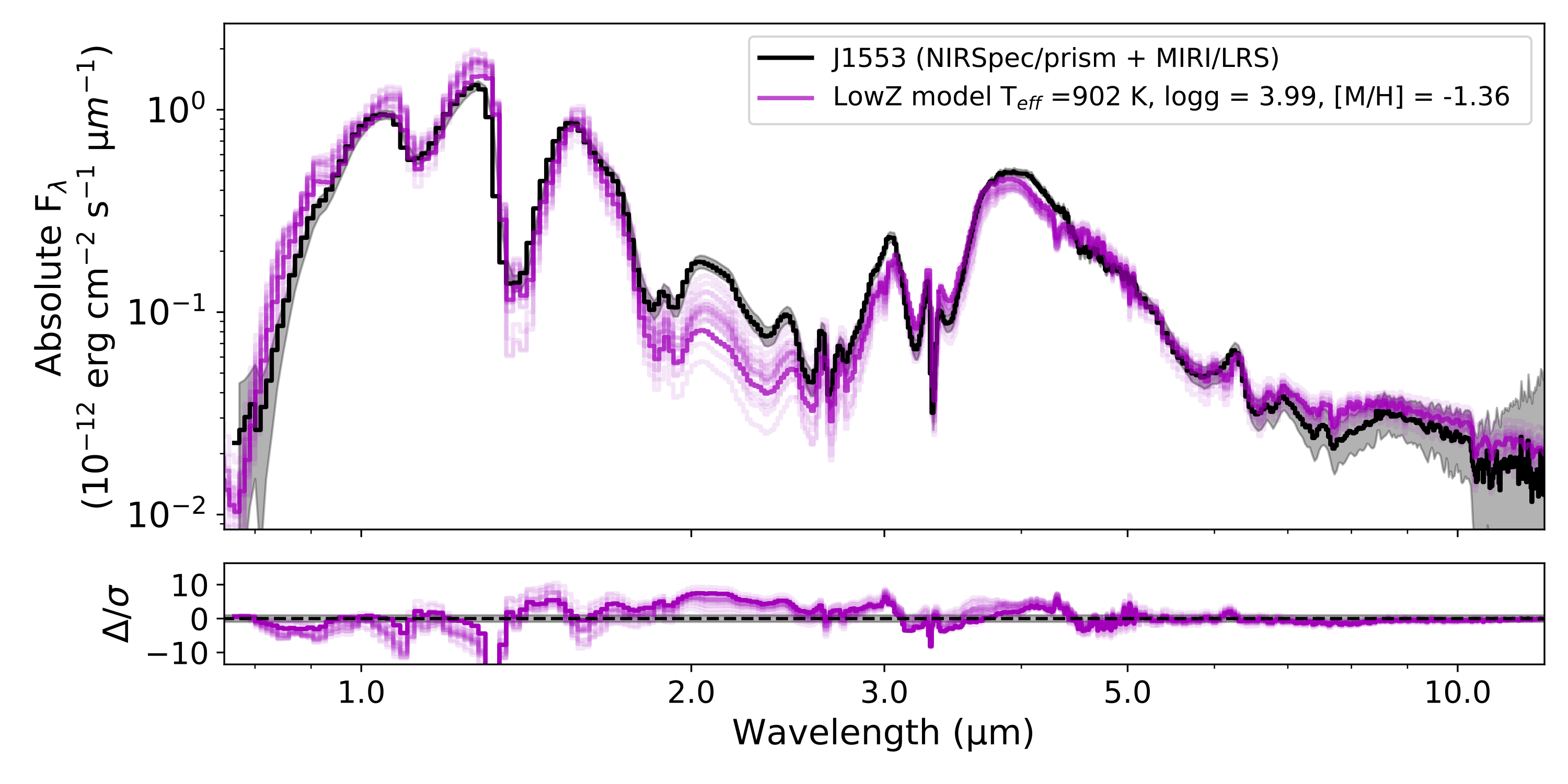} \\
\includegraphics[width=0.75\textwidth]{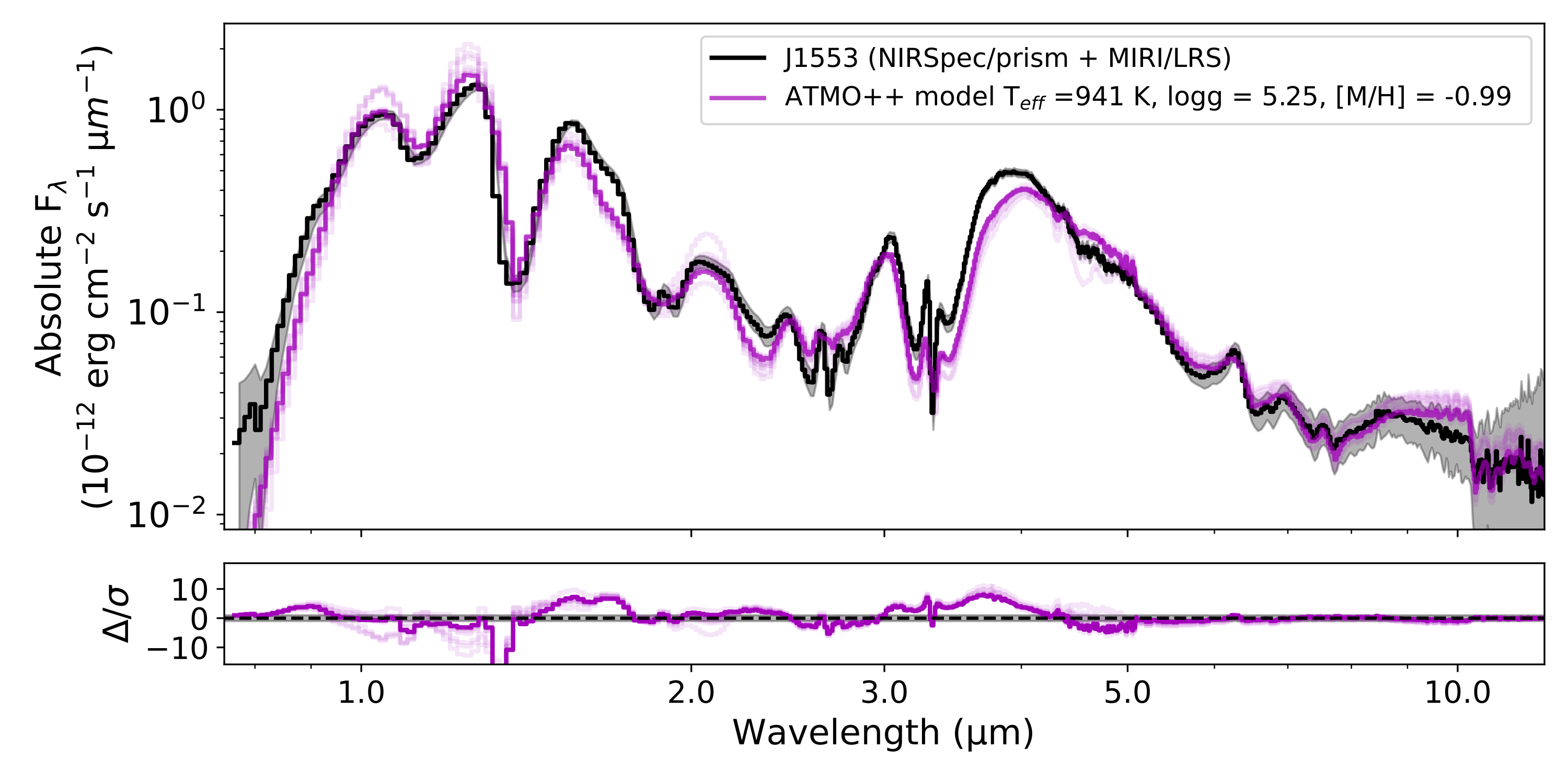}
\caption{Comparison of the absolute flux-calibrated JWST NIRSpec/prism plus MIRI/LRS low-resolution spectrum (black line) to the best-fitting 
SAND (top; \citealt{Alvarado_2024,Gerasimov_2024a}),
LowZ (middle; \citealt{Meisner_2021}), and
ATMO++ (bottom; \citealt{Phillips_2020}) models (magenta lines).
Thick magenta lines show the best overall fit, thin semi-transparent lines show posterior draws from the MCMC chain.
The bottom portion of each panel shows the residuals (data minus model) relative to uncertainty ($\Delta/\sigma$), assuming a factor of 10 increase in the flux density uncertainties (gray shaded region in top panels; see discussion in $\S$\ref{sec:grid}). SAND provides the best overall fit among the model grids considered.
\label{fig:modelfit}}
\end{figure*}

The three best-fitting models were 
SAND ($X^2$ = 3.2),
LowZ ($X^2$ = 7.6), and
ATMO++ ($X^2$ = 9.6),
each yielding 900~K $<$ {\teff} $<$ 1000~K and $-$1.5~dex $\leq$ [M/H] $\lesssim$ $-$1~dex.
To refine these values, we conducted a Markov Chain Monte Carlo (MCMC) fit to the J1553 spectrum with \texttt{ucdmcmc} following the approach described in 
\cite{Burgasser_2025,Burgasser_PH3} and \cite{2026AJ....171..191M}. In brief, we followed a Metropolis-Hastings algorithm \citep{1953JChPh..21.1087M,HASTINGS01041970}, using a single chain of 10,000 steps to sample model parameters within each grid. Models were linearly interpolated in logarithmic flux units on a continuous logarithmic parameter scale ({\teff} $\Rightarrow$ $\log${\teff}). At each iteration, new parameters were drawn based on normal distributions centered at the current parameter set (initially the best-fit grid parameters), and the fit for the new interpolated model ($i+1$) was compared to the fit for the prior model ($i$) using a $X^2$-based statistic
\begin{equation} \label{eqn:acceptance}
    \frac{X^2(i+1)-X^2(i)}{MIN[X^2]} < \mathcal{U}(0,0.5)
\end{equation}
where $MIN[X^2]$ is the minimum of all $X^2$ values in the chain and 
$\mathcal{U}(0,0.5)$ is a number drawn from a uniform distribution between 0 and 0.5. If the new fit satisfied this criterion, these parameters were added to the chain; otherwise, the previous parameters were added.
As discussed in \cite{ZJ_Zhang_2021}, the potential for underestimated physical parameter uncertainties is a concern when fitting atmospheric models to observed brown dwarf spectra, driven by both systematic errors in the models and underestimation of the measured spectral flux density uncertainties. The acceptance statistic defined in Equation~\ref{eqn:acceptance} minimizes the influence of the latter issue by normalizing to the minimum $X^2$. We validated this characteristic by conducting MCMC fits with the SAND models using the original flux density uncertainties and scaling the uncertainties by factors of 5, 10 and 15. All four fits yielded equivalent parameter values and uncertainties, with $X^2$ values scaling inversely with the square of the uncertainty scaling factor. We deduced an optimal uncertainty scaling of 10 to enforce $X^2 \approx 3$ for the SAND models, and used this scaling for all of the MCMC model fits.
The resulting parameter values and uncertainties, computed from the median and 16\% and 84\% quantiles ($\pm$1$\sigma$) of the chain distributions, are reported in Table~\ref{tab:modelfit}. 
Figure~\ref{fig:modelfit} displays the best-fit (lowest $X^2$) model spectrum and random draws from the parameter chains for the three model sets.

\begin{deluxetable}{lccc}
\tabletypesize{\small}
\tablewidth{0pt} 
\tablecaption{MCMC Spectral Model Fit Analysis \label{tab:modelfit}}  
\tablehead{
\colhead{Parameter} & \colhead{SAND} & \colhead{LowZ} & \colhead{ATMO++}  
}
\startdata 
$T_{\rm eff}$ (K) &  960$^{+28}_{-32}$  & 917$^{+62}_{-46}$ & 917$^{+63}_{-51}$     \\ 
$\log{g}$ (cm~s$^{-2}$) &  4.53$^{+0.30}_{-0.16}$  &  3.95$^{+0.44}_{-0.34}$ & 4.88$^{+0.32}_{-0.79}$  \\
{[M/H]} (dex) &  $-$1.16$^{+0.28}_{-0.25}$  & $-$1.48$^{+0.17}_{-0.20}$ & $-$0.82$^{+0.19}_{-0.12}$    \\
{[$\alpha$/Fe]} (dex) &  +0.15$^{+0.04}_{-0.06}$   &  \nodata &  \nodata   \\
C/O &  \nodata  &  0.15$^{+0.10}_{-0.05}$  &  \nodata    \\
$\log\kappa_{zz}$ (cm$^2$~s$^{-1}$) &  \nodata  &  1.1$^{+0.6}_{-1.1}$ & 4.7$^{+0.9}_{-0.5}$   \\
Radius\tablenotemark{a} (R$_{Jup}$) &  0.78$\pm$0.10   &  0.79$\pm$0.13 & 0.83$\pm$0.13    \\
Min $X^2$\tablenotemark{b} &  3.2  & 7.6  &  9.6   \\
\enddata
\tablenotetext{a}{Radius based on optimal scaling factor $\alpha$ when adopting the photometric distance of $39^{+5}_{-4}$~pc and accounting for the distance uncertainty.}
\tablenotetext{b}{Minimum $X^2$ accounts for a factor of 10 increase in the spectral flux uncertainty; see discussion in $\S$\ref{sec:grid}.}
\end{deluxetable}

The best-fit SAND model is a reasonable fit to the data 
with modest departures ($\lesssim$10-20\%) primarily in the 1--1.8~$\mu$m range and underestimated CH$_4$ opacity at 3.3~$\mu$m.
The LowZ model shows larger departures in the 2--3~$\mu$m continuum, while the ATMO++ models shows major deviations in the 1.6, 2.2, and 3.3~$\mu$m CH$_4$ bands.  All three model sets yield statistically consistent effective temperatures, with the SAND
$T_{\rm eff}$ = 960$^{+28}_{-32}$~K
in good agreement with the temperature derived from \Lbol~when adopting the photometric distance ($\S$\ref{sec:lbol}).
For metallicity, all three models predict a consistently subsolar metallicity, 
with the SAND models yielding $-1.16^{+0.28}_{-0.25}$~dex, more characteristic of the Milky Way's (inner) halo rather than its thin or thick disk \citep{Carollo_2007}. The LowZ model yields an even lower metallicity of $-1.48^{+0.17}_{-0.20}$~dex, albeit consistent with the SAND model given the uncertainties, while fits to the ATMO++ models run into the low-metallicity limit of that set.
The SAND models further indicate modest alpha element enrichment, [$\alpha$/Fe] = +0.15$^{+0.04}_{-0.06}$~dex, a common feature for metal-poor ``low-alpha'' halo stars which may have been accreted from dwarf satellite galaxies
\citep{2010A&A...511L..10N,Mackereth_2019,2024AJ....167....6R}.
The LowZ grid also indicates a low C/O = 0.15$^{+0.10}_{-0.05}$, which is consistent with the C/O = $0.20 \pm 0.06$ for Milky Way stars with [Fe/H] $< -1$~dex \citep{2004A&A...414..931A}.
Finally, we find considerable variance among surface gravities from these fits. 
The SAND-derived {\logg} = 4.53$^{+0.30}_{-0.16}$ (cgs) is somewhat lower than the \logg~$\approx$ 5.0--5.5~(cgs) range expected for an old brown dwarf \citep{Burrows_2001},
but consistent with the {\logg} = 4.78$^{+0.35}_{-0.20}$ (cgs) from prior near-infrared fits \citep{Burgasser_2025}.
The ATMO++ {\logg} = 4.88$^{+0.32}_{-0.79}$ (cgs) is more consistent with expectations of a high gravity thick disk or halo brown dwarf, albeit with large uncertainties; while 
the LowZ \logg~= 3.95$^{+0.44}_{-0.34}$ (cgs) is quite low and typical of a very young (age $\lesssim$ 10~Myr) brown dwarf.
We note that low surface gravities have also been reported in fits to NIRSpec/prism spectra
of distant brown dwarfs in deep JWST surveys \citep{2025ApJS..281...49T}, suggesting a systematic bias for these low-resolution data.
Indeed, the lowest (highest) surface gravities in these fits come from the models with the lowest (highest) metallicities, suggesting a correlated trend likely tied to pressure-sensitive features such as CIA H$_2$.

\subsection{Radial Velocity} \label{sec:kinematics}

To determine a 3D velocity for J1553, we measured its RV from the NIRSpec/G395H spectrum using
the Spectral Modeling Analysis and RV Tool (SMART) software package \citep{Dino_smart_zenodo,Dino_smart_paper}.
The full description of our fitting workflow is detailed in \cite{Dino_smart_paper}.
In short, we forward model the CO 1-0 fundamental band in the 4.4--5.0~$\mu$m range, as the oscillatory structure of this feature provides a robust anchor of radial motion (cf.\ RV inference from the CO 2-0 first overtone band at $\sim$2.3~{\micron}; \citealt{Blake_2010, Hsu_2023}).
We used the high-resolution Sonora Elf Owl models \citep{Mukherjee_2024} as our model template, interpolated as a function of effective temperature, surface gravity, metallicity, vertical eddy diffusion parameter ($\kappa_\mathrm{zz}$), and C/O ratio. These parameters
are fit simultaneously with rotational velocity broadening ({\vsini}), RV, and instrumental line spread function kernel, the last modeled as a Gaussian function.
We also included a second-order polynomial fit to the continuum, as we are primarily interested in matching the structure of the CO band.
The best-fit parameters were obtained using the MCMC sampling method \texttt{emcee} \citep{Foreman-Mackey_2013} with 100 walkers and 2000 steps, removing the first 1000 steps as burn-in.

\begin{figure}
\centering
\includegraphics[width=0.8\textwidth]{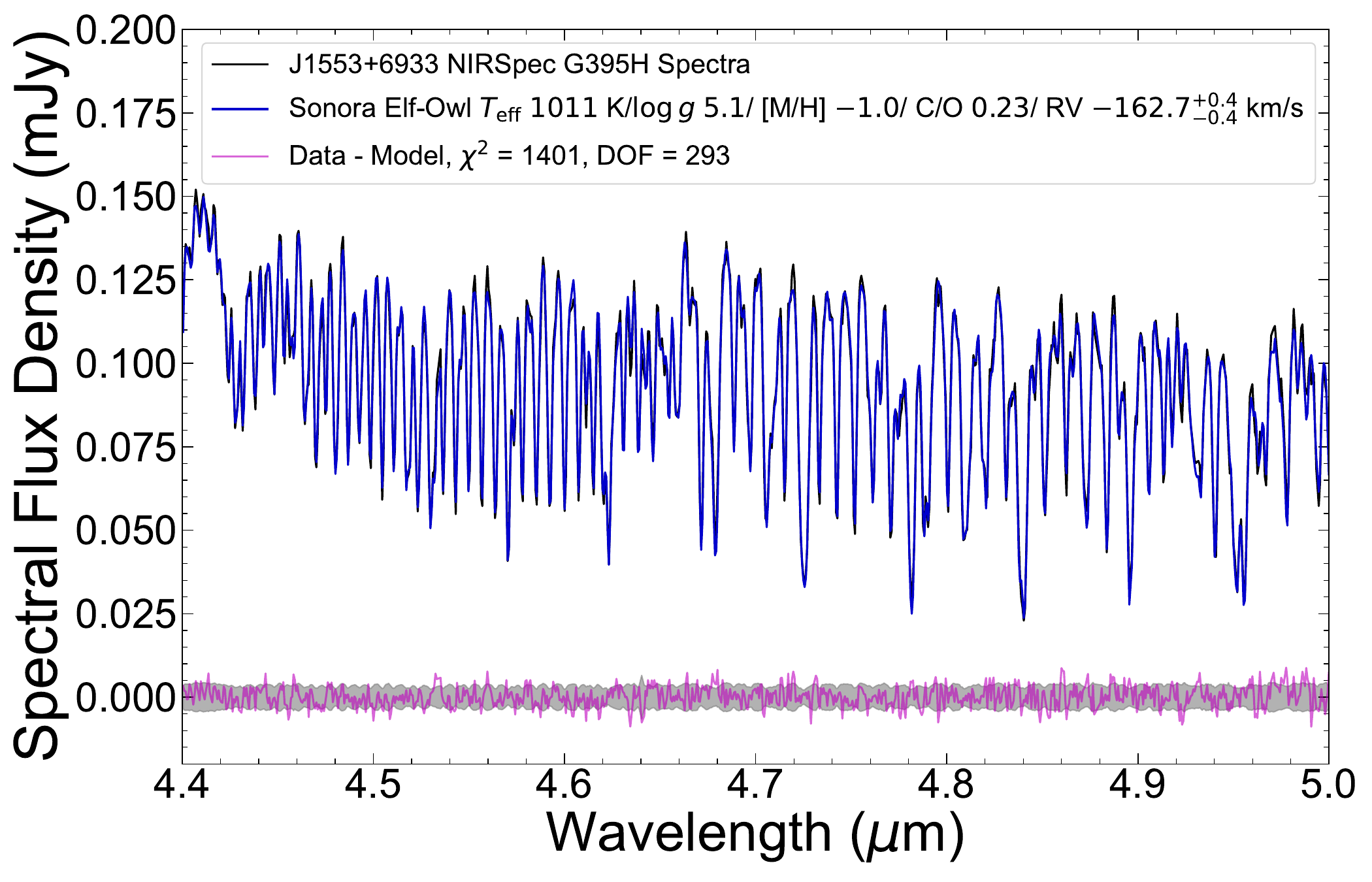}
\caption{Forward model fit to the 4.4--5.0~$\mu$m NIRSpec/G395H spectrum of J1553 based on SMART forward modeling analysis \citep{Dino_smart_paper,Dino_smart_zenodo}.
The observed spectrum is plotted in black and the best-fit Sonora Elf Owl model \citep{Mukherjee_2024} is plotted in blue.
The residuals (data $-$ model) are depicted in magenta, and compared to the $\pm$1$\sigma$ spectral data uncertainty shown by the gray shaded region.
Fit parameters are listed in the legend.
\label{fig:rv_fit}}
\end{figure}

The resulting best-fit model is illustrated in Figure \ref{fig:rv_fit}.
The best-fit model matches well with the observed G395H spectra, demonstrating the reliability of using this band to derive RV.
We find a best-fit radial velocity for J1553 of $-162.7 \pm $0.4 km/s, where this uncertainty reflects only the statistical uncertainty of our MCMC analysis.  We include an additional 
$\pm 5$~km/s RV uncertainty to account for
systematic error based on observations of sources with independent high-resolution spectroscopic RVs \citep{2025jwst.rept.9239G,Hsu_2026}. 
This RV differs substantially from the +110 $\pm$ 90~km/s reported by \citet{Burgasser_2025} based on ground-based data,
the large uncertainty in that measurement due to the lack of strong features in the near-infrared spectrum analyzed.

We note that the best-fit Sonora Elf Owl model to the G395H spectrum has \teff~= 1011~K and [M/H] = $-$1~dex, at the edge of the model grid, both in reasonable agreement with the parameters inferred from the low-resolution spectral fit presented in $\S$\ref{sec:grid} (Table \ref{tab:modelfit}). Similarly, the low C/O = 0.23 favored by the G395H fit is consistent with the C/O = 0.15$^{+0.10}_{-0.05}$ derived from the LowZ model fit to the low-resolution data.
However, we caution against over-interpreting the stellar atmospheric parameters from the G395H fit -- given the relatively narrow spectral range and fitting out of the continuum, concerns have been previously raised in high-resolution spectroscopic modeling \citep{Del_Burgo_2009, Dino_smart_paper}. 
Nevertheless, the consistency in physical parameters across different models and spectral resolutions indicates these parameters are likely robust.

\section{Discussion: Is J1553 a Member of Gaia-Enceladus?} \label{sec:discussion}

Gaia-Enceladus, also referred to as the Gaia Sausage or Gaia-Sausage-Enceladus, is the debris of a major Galactic merger event that  occurred $\approx 8$–11 Gyr ago \citep{Helmi_2018,Belokurov_2018}. The progenitor was a massive dwarf galaxy whose stellar members are now a prominent component of the Milky Way's inner halo. Gaia-Enceladus members are generally selected based on their eccentric, often retrograde orbits with low azimuthal actions \citep[e.g.,][]{Myeong_2019,Naidu_2020,Horta_2023,Feuillet_2021}, as well as ``low'' alpha element enrichment \citep[e.g.,][]{Mackereth_2019}. Because they are widely distributed through the Milky Way, it is possible to find Gaia-Enceladus members in the solar neighborhood \citep[e.g.,][]{Kim_2022} if they happen to be passing through the disk close to the Sun. 

Here, we assess the evidence that J1553 is a brown dwarf member of Gaia-Enceladus based on its kinematics and chemical composition.

\subsection{J1553 Kinematics in Relation to Gaia-Enceladus}
\label{sec:kinematics_discussion}

\begin{figure*}
\includegraphics[width=7.0in]{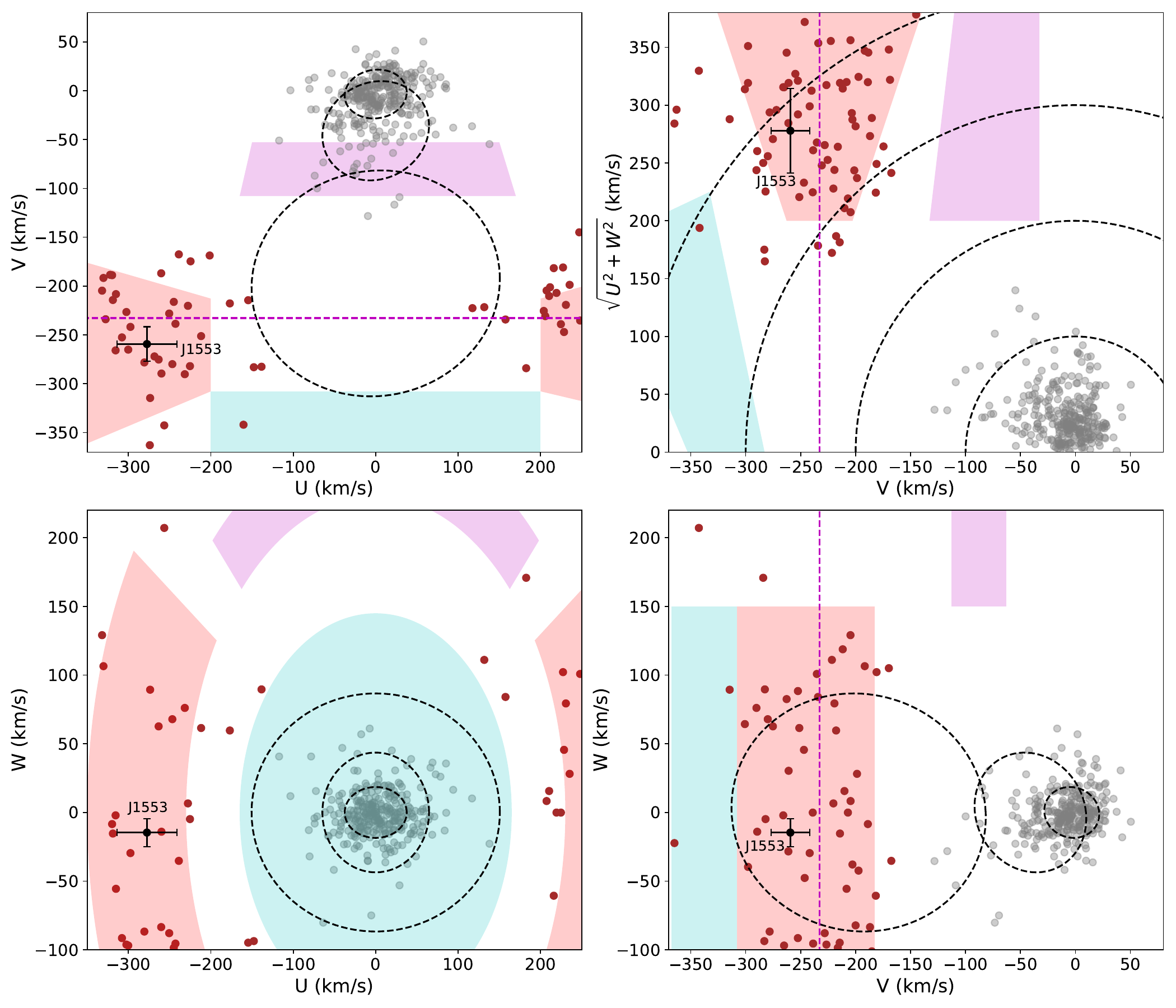}
\caption{Kinematics of J1553 in the local standard of rest. 
Top left: $V$ versus $U$;
bottom left: $W$ versus $U$;
bottom right: $W$ versus $V$;
top right: Toomre diagram ($\sqrt{U^2+W^2}$ versus $V$).
In each panel, we show J1553's location using our JWST-measured RV, the proper motion from \cite{Zhang_2025}, and the photometric distance from \citet{Meisner_2020b}.
Gray dots show a random subset of 300 stars from the Gaia Catalog of Nearby Stars \citep{GCNS} with RV uncertainty $ < 5$~km/s, a sample dominated by thin disk stars.
Brown dots show kinematically-selected Gaia-Enceladus members from \cite{Ernandes_2024}.
Colored regions highlight velocity spaces occupied by 
Gaia-Enceladus \citep[light red;][]{Belokurov_2018,Helmi_2018},
Thamnos 2 \citep[light cyan;][]{Koppelman_2019}, 
and Helmi streams \citep[light magenta;][]{Helmi_1999}, based on \cite{Koppelman_2019}. 
In the paired velocity panels, dashed black lines indicate the 1$\sigma$ velocity dispersion spheres for chemically-selected thin disk, thick disk, and halo stars (in order of increasing size) based on Gaia DR2 \citep{GAIA} and APOGEE \citep{Anguiano_2020} measurements. 
In the Toomre diagram, dashed black circles represent contours of total LSR speed $\sqrt{U^2 + V^2 + W^2}  = 100, 200, 300, 400$~km/s. 
In all panels plotting $V$ velocity, the dashed magenta line separates prograde and retrograde orbits based on a local circular speed $V_{circ}$ = $-$230~km/s \citep{McMillan_2017}. All error bars shown for J1553 are 1$\sigma$.
\label{fig:uvw}}
\end{figure*}

Figure \ref{fig:uvw} shows the $UVW$ space velocities for J1553 in the local standard of rest (LSR)\footnote{We adopt a right-handed coordinate system with $U$ pointed radially inward, $V$ pointed in the direction of Galactic rotation, and $W$ pointed toward the North Galactic Pole. To correct to the LSR frame, we adopt a solar motion of ($U_\odot,V_\odot,W_\odot$) = (11.1,12.24,7.25)~km/s from \citet{Schonrich_2010}.} incorporating our JWST-measured RV and the proper motion reported in \citealt{Zhang_2025} (Table \ref{tab:data}).
We computed velocities using the photometric distance estimate of \cite{Meisner_2020b}.
J1553's motion far exceeds those of local thin disk stars \citep{GCNS}, and comparison to population kinematics indicates it is far more likely to be a halo brown dwarf than a thick disk brown dwarf, with $p$(halo)/$p$(thick disk) = $1.4 \times 10^{7}$ \citep{Bensby_2003}. J1553 lies near the prograde/retrograde boundary ($V_{circ} \approx$ $-$230~km/s; \citealt{McMillan_2017}) and is retrograde when adopting its photometric distance estimate.
The small $L_{z}$ angular momentum indicated by J1553's proximity to this boundary implies a highly eccentric Galactic orbit.
Using \verb|galpy| \citep{galpy} and its MWPotential2014 Galactic potential, we find a high eccentricity of 0.94$\pm$0.03 for J1553.

Previous studies have noted that halo stars with high eccentricity and low angular momentum orbits
are members of Gaia-Enceladus \citep{Belokurov_2018,Mackereth_2019,Feuillet_2021,Horta_2023}.
Figure~\ref{fig:uvw} shows the approximate velocity loci of 
Gaia-Enceladus \citep{Belokurov_2018,Helmi_2018},
Thamnos 2 \citep{Koppelman_2019}, 
and Helmi streams \citep{Helmi_1999} based on \cite{Koppelman_2019}. 
J1553 lies squarely within the Gaia-Enceladus regions for every two-parameter combination of $UVW$ velocity, as well as in the Toomre diagram.
We also show the velocities of 73 kinematically-selected Gaia-Enceladus members from \cite{Ernandes_2024}, which show some scatter beyond the \cite{Koppelman_2019} boundaries. Nevertheless, J1553's kinematic alignment with this Gaia-Enceladus sample is compelling.

For completeness, \cite{Burgasser_2025} identified J1553 as a potential kinematic match to the Helmi streams \citep{Helmi_1999}. With the more accurate and precise RV from our JWST measurements, this association can be firmly ruled out.

\subsection{J1553 Chemical Composition in Relation to Gaia-Enceladus}

\begin{figure*}
\centering
\includegraphics[width=5in]{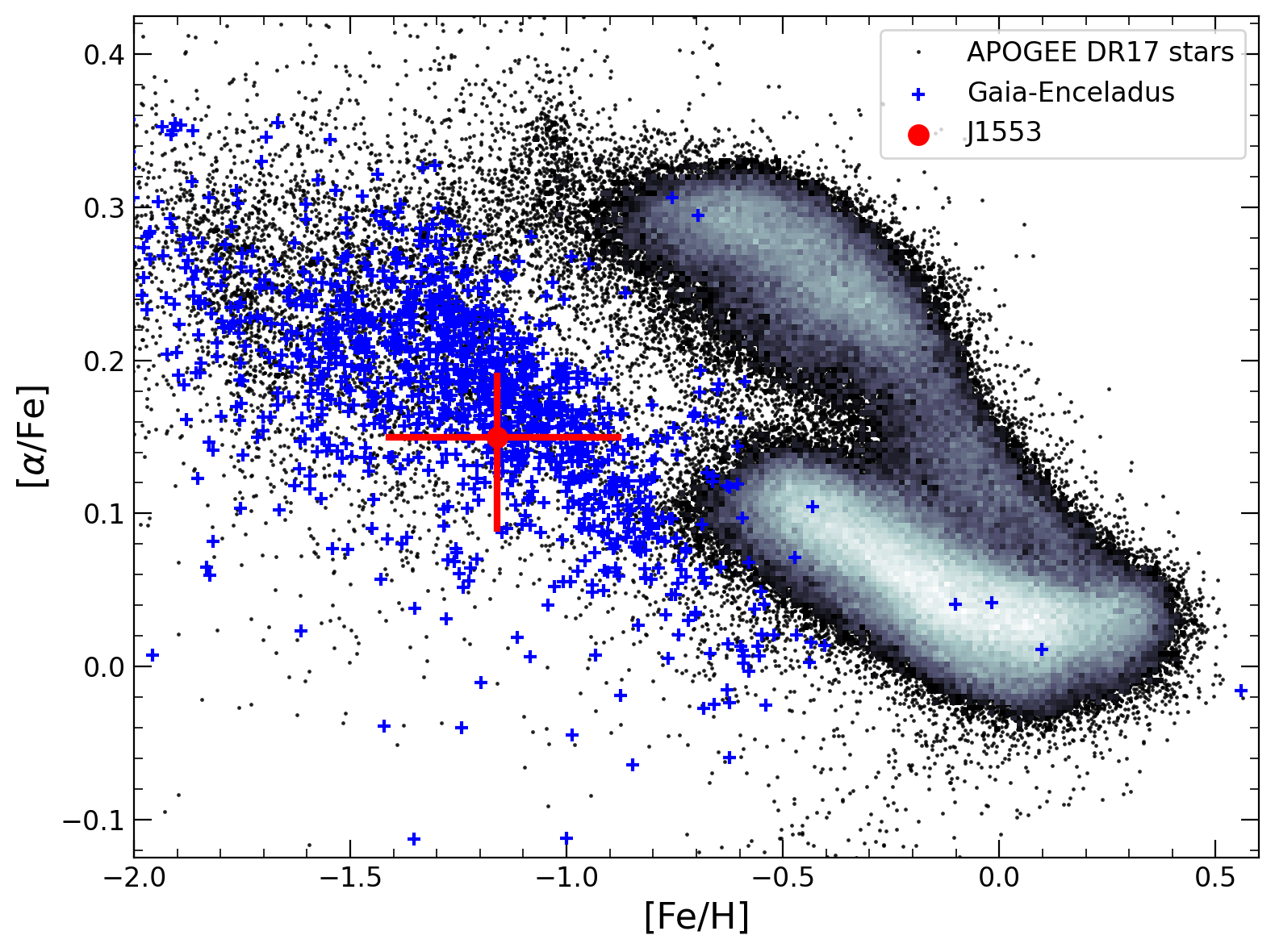}
\caption{Trend of [$\alpha$/Fe] versus [Fe/H] for Milky Way stars patterned after Figure 6 of \cite{Horta_2023}. The gray density map and small black data points represent a sample of $\approx 170,000$ APOGEE DR17 \citep{DR17} stars selected similarly to the `parent sample' of \cite{Horta_2023}. The blue plus marks are Gaia-Enceladus members drawn from this APOGEE parent sample, selected based on the criteria $|L_z| < 500$~kpc km/s, $-1.6 \times 10^{5}$~km$^2$/s$^2 < E_{tot} < -1.1 \times 10^{5}$~km$^2$/s$^2$ (using the \citealt{McMillan_2017} Milky Way potential), and [Al/Fe] $< 0.1$~dex. The red data point represents the J1553 abundances inferred from our SAND model fits.
\label{fig:abundances}}
\end{figure*}

The metallicity and alpha element enrichment values inferred from spectral model fits to J1553 are consistent with the ``low-alpha'' halo, suggesting an accretion origin. Prior studies have measured the metallicity distribution function (MDF) of Gaia-Enceladus stars, with mean values in the range $-1.45 \ \textrm{dex} \lesssim \textrm{[Fe/H]} \lesssim -1.10$~dex \citep{Feuillet_2020,Feuillet_2021,Naidu_2020,Bonifacio_2021,Buder_2022}. A ``consensus'' MDF with $\langle$[Fe/H]$\rangle$ = $-$1.2~dex and $\sigma_{\rm [Fe/H]} = 0.3$~dex is fully consistent with the [M/H] = $-1.16^{+0.28}_{-0.25}$~dex inferred from the SAND model fits.
Gaia-Enceladus stars also show a trend of increasing [$\alpha$/Fe] with decreasing metallicity \citep[e.g., Figure 2 of][]{Helmi_2018}. Near [Fe/H] = $-1.15$~dex, the \cite{Helmi_2018} Gaia-Enceladus stars span $0 \ \textrm{dex} \lesssim$ [$\alpha$/Fe] $\lesssim$ 0.45~dex, with the bulk at $0.15 \ \textrm{dex} \lesssim$ [$\alpha$/Fe] $\lesssim$ 0.3~dex. 
Other studies have found similar levels of enrichment in alpha tracers such as [Mg/Fe]; e.g., $0.15 \ \textrm{dex} \lesssim$ [Mg/Fe] $\lesssim$ 0.3~dex from \citet{Feuillet_2021}, and 0.1~dex $\lesssim$ [Mg/Fe] $\lesssim$ 0.3~dex from \citet{Horta_2023}.
Again, these values are fully consistent with the [$\alpha$/Fe] = +0.15$^{+0.04}_{-0.06}$~dex inferred from our SAND model fits. Figure \ref{fig:abundances} illustrates the match between our inferred J1553 [$\alpha$/Fe] and measurements of [$\alpha$/Fe] for a large sample of Gaia-Enceladus members from APOGEE \citep{APOGEE_canonical}. 

In summary, J1553's kinematics, overall bulk metallicity, and alpha enrichment are all consistent with Gaia-Enceladus membership, making this the first brown dwarf to have multiple lines of evidence associating it with this accreted Milky Way structure.

\begin{deluxetable}{lccc}
\tablecaption{Luminosity, Kinematics, and Estimated Mass of WISEA J155349.96+693355.2 \label{tab:data}}
\tablehead{
\colhead{Parameter} & \colhead{Value} & \colhead{Ref.}}
\startdata
\multicolumn{3}{c}{Observed Properties} \\
\hline
RV  (km/s) & $-163$ $\pm$ 5 & 1\\
$\mu_\alpha\cos\delta$  (mas/yr) & $-1527 \pm 12$ & 2\\
$\mu_\delta$  (mas/yr) & $+1258 \pm 13$ & 2\\
$d_\mathrm{phot}$ (pc) & 39$^{+5}_{-4}$& 3\\
\hline
\multicolumn{3}{c}{Inferred Properties} \\
\hline
log(\Lbol/$L_{\odot}$) & $-$5.29 $\pm$ 0.10 & 1 \\
$M$ (\mjup) & 67$_{-4}^{+2}$  & 1 \\
U (km/s) & $-280 \pm 41$ & 1 \\
V (km/s) & $-261 \pm 20$ & 1 \\
W (km/s) & $-14 \pm 11$ & 1 \\
\enddata
\tablerefs{
(1) This work;
(2) \citet{Zhang_2025};
(3) \citet{Meisner_2020b}.}
\tablecomments{Inferred properties assume the photometric distance. Galactic $UVW$ velocities quoted here have been corrected to the LSR following \cite{Schonrich_2010}. Mass has been computed using the \cite{Saumon_2008} evolutionary models and assuming an age of $10 \pm 3$~Gyr.}
\end{deluxetable}

\section{Conclusion} \label{sec:conclusion}

We have presented JWST NIRSpec and MIRI 0.6-12~$\mu$m spectral observations of the T subdwarf J1553, which verify its low-temperature, metal-poor, and substellar nature.
In particular, we find evidence of enhanced CIA H$_2$ absorption and weakened molecular gas features, including an absence of CO$_2$, consistent with a metallicity intermediate between 
Wolf~1130C ([M/H] = $-0.68 \pm 0.04$~dex; \citealt{Burgasser_PH3})
and J1534 ([M/H] = $-2.22 \pm 0.05$~dex; \citealt{Faherty_silane}).
Analysis of its integrated bolometric flux and spectral model fits indicate a temperature discrepancy that is resolved if its actual distance is larger than that based on the low S/N parallax measurement of \cite{Zhang_2025}. The photometric distance of $\approx 39$~pc, originally estimated by \citet{Meisner_2020b}, yields a consensus temperature {\teff} $\approx$ 950~K and a luminosity {\Lbol} = $-$5.29$\pm$0.10.
Our JWST data also allow us to measure an RV = $-163 \pm 5$~km/s, which combined with the \cite{Zhang_2025} proper motion measurement and photometric distance indicates that J1553 has a high eccentricity Galactic orbit.
Kinematics associated with the photometric distance align precisely with known members of the Gaia-Enceladus structure, and this association is strengthened by J1553's bulk metallicity and modest alpha element enhancement.
Overall, we find this source to be the first compelling candidate to be an accreted brown dwarf. 

To validate these findings, future work is needed to obtain a better distance measurement for J1553 to resolve the discrepancy between current astrometric and photometric estimates, and to improve the precision of $UVW$ velocity components. With J $\approx$ 19.1 mag (Vega), additional J1553 J-band astrometry could be obtained from the ground. We expect that a continuation of the \cite{Zhang_2025} J1553 ground-based J-band astrometric monitoring campaign with a longer time baseline 
would yield an improved trigonometric parallax.
In addition, our characterization of the atmosphere of J1553 can be improved through 
more precise retrieval modeling of the G395H spectrum, which can provide individual elemental abundances as traced by key molecular features (e.g., H$_2$O, CH$_4$, CO, CO$_2$, NH$_3$, H$_2$S), as well as search for evidence of trace molecules such as PH$_3$, SiH$_4$, and isotopologues.
As the number of metal-poor brown dwarfs with extreme Galactic orbits grows (e.g., \citealt{2019MNRAS.486.1840Z,2024MNRAS.533.1654Z,2024ApJ...971L..25B,Burgasser_2025,2025arXiv251206069A}), 
both infrared astrometric campaigns and high-sensitivity infrared spectra will be needed to obtain the velocity and compositional information to explore their possible extragalactic origins.

\begin{acknowledgments}

JWST data used in this work are available at MAST: \dataset[doi: 10.17909/qe1k-pe44]{\doi{10.17909/qe1k-pe44}}. We thank the anonymous referees. AB, CCH, and CAT acknowledge funding 
from NASA/STScI through grant JWST-GO-04668.007, and  from the Heising-Simons Foundation. NL acknowledge support from the Agencia  Estatal de Investigaci\'on del Ministerio de Ciencia e Innovaci\'on (AEI-MCINN) under grant PID2022-137241NB-C41\@. This work is based (in part) on observations made with the NASA/ESA/CSA James Webb Space Telescope. The data were obtained from the Mikulski Archive for Space Telescopes at the Space Telescope Science Institute, which is operated by the Association of Universities for Research in Astronomy, Inc., under NASA contract NAS 5-03127 for JWST. These observations are associated with program JWST-GO-04668. AM acknowledges support from JWST-GO-03558 and JWST-GO-6084.
\end{acknowledgments}

\begin{contribution}

A. Meisner led the write-up of this manuscript and assessment of Gaia-Enceladus association. 
A. Burgasser is PI of JWST-GO-04668,
led the model grid fits, and contributed to the analysis. 
C.-C. Hsu led the radial velocity measurement, calculation of space velocities, and assessment of full kinematics. 
S. Alejandro Merchan and J. Faherty led the bolometric luminosity analysis and comparison against literature JWST old/cold brown dwarf spectra. G. Su\'arez provided absorption cross-sections.

\end{contribution}

\facilities{JWST(NIRSpec, MIRI)}

\software{
astropy \citep{astropy:2013,astropy:2018,2022ApJ...935..167A},
\texttt{galpy} \citep{galpy},
JWST Calibration Pipeline \citep{Bushouse_2024}, 
SEDkit \citep{SEDkit}, 
SMART \citep{Dino_smart_zenodo,Dino_smart_paper}, 
SPLAT \citep{SPLAT}, 
\texttt{ucdmcmc} \citep{ucdmcmc}.
}

\bibliography{sample7}{}
\bibliographystyle{aasjournalv7}

\end{document}